\documentstyle[12pt]{article}
\input epsf
\def \b{{\cal B}}

\def \beq{\begin{equation}}
\def \eeq{\end{equation}}
\def \beqn{\begin{eqnarray}}
\def \eeqn{\end{eqnarray}}
\def \cn{Collaboration}

\def \ite{{\it et al.}}

\def \s{\sqrt{2}}

\def \st{\sqrt{3}}
\def \sx{\sqrt{6}}
\def \v#1#2{V_{#1#2}}

\begin{document}
\renewcommand{\thetable}{\Roman{table}}
\rightline{EFI-99-20}
\rightline{hep-ph/9905366}
\rightline{May 1999}
\bigskip
\bigskip
\centerline{\bf FINAL-STATE PHASES IN CHARMED MESON}
\centerline{{\bf  TWO-BODY NONLEPTONIC DECAYS}
\footnote{To be submitted to Phys.~Rev.~D.}}
\bigskip
\centerline{\it Jonathan L. Rosner}
\centerline{\it Enrico Fermi Institute and Department of Physics}
\centerline{\it University of Chicago, Chicago, IL 60637}
\bigskip
\centerline{\bf ABSTRACT}
\medskip
\begin{quote}

Observed decay rates indicate large phase differences among the amplitudes for
the charge states in $D \to \bar K \pi$ and $D \to \bar K^* \pi$ but relatively
real amplitudes in the charge states for $D \to \bar K \rho$. This feature is
traced using an SU(3) flavor analysis to a sign flip in the contribution of one
of the amplitudes contributing to the latter processes in comparison with its
contribution to the other two sets.  This amplitude may be regarded as an
effect of rescattering and is found to be of magnitude comparable to others
contributing to charmed particle two-body nonleptonic decays.
\end{quote}
\medskip
\leftline{\qquad PACS codes:  13.25.Ft, 11.30.Hv, 14.40Lb}
\bigskip

\centerline{\bf I. INTRODUCTION}
\bigskip

The phases of amplitudes in weak two-body nonleptonic decays of heavy mesons
are of interest in the search for CP violation.  Decays of $B$ mesons in many
cases are expected to receive contributions from more than one weak subprocess.
If the corresponding amplitudes also differ in their strong phases one can
expect to see CP-violating asymmetries \cite{CSBTP} in $B$ meson decay rates. 
The origin and magnitude of such strong phase differences has been the subject
of much discussion \cite{resc,JRFSI}. 

The nonleptonic decays of charmed mesons are expected to involve weak
amplitudes with very similar phases \cite{KM}. If decay amplitudes exhibit
large phase differences they are almost certainly due to strong final-state
interactions.  Thus these decays can serve as a laboratory for the examination
of final-state effects.  The lower mass of charmed particles in comparison with
$B$ mesons tends to amplify these effects, which are expected to diminish in
relative importance with increasing energy. 

One class of charmed meson decays in which final-state interactions can be
well-probed experimentally is the set of Cabibbo-favored nonleptonic two-body
decays governed by the subprocess $c \to s u \bar d$.  By comparing decay
rates, one finds that the amplitudes for the three charge states $\bar K^0
\pi^+$, $K^- \pi^+$, $\bar K^0 \pi^0$ in $D \to \bar K \pi$ decays cannot all
be real with respect to one another \cite{MkIII,Kamal,Banff,SuzDB}.  A similar
conclusion can be drawn both from decay rates and from relative phases of
Dalitz plot amplitudes for the three charge states in $D \to \bar K^* \pi$
\cite{MkIII,Anjos,ARGUS,Frab}.  However, both decay rates and Dalitz plot
analyses reveal no relative phases between the amplitudes in the three charge
states of $D \to \bar K \rho$. 

In the present paper we examine the source of this apparent difference between
$\bar K \pi, \bar K^* \pi$ decays on the one hand and $\bar K \rho$ decays on
the other.  We find that effects of strong final-state interactions are present
in all three decays, but they contribute to the various charge states in $\bar
K \rho$ decays in such a way that the amplitudes are all relatively real.  This
result has some implications for the universality of strong
final-state-interaction effects. In passing, we note some simple regularities
of contributions to these processes which can be related to those in
semileptonic decays. 

We consider only Cabibbo-favored decays, in order to focus on the comparison
between $\bar K \pi, \bar K^* \pi$, and $\bar K \rho$.  The framework we
employ is an SU(3) flavor analysis \cite{Quigg,Zepp,SW} which can be expressed
in terms of quark graphs \cite{Chau,GHLR} but whose interpretation in those
terms should not be taken too literally. We assume the following contributions:
 (1) A color-favored ``tree'' amplitude $T$, (2) a ``color-suppressed'' tree
amplitude $C$, (3) an ``exchange'' amplitude $E$ contributing only to $D^0$
decays, and (4) an ``annihilation'' amplitude $A$ contributing only to $D_s$
decays.  The amplitudes $T$ and $C$ do not involve the spectator quark, while
$E$ and $A$ do.  They are most likely parametrizations of rescattering effects,
since when taken literally as short-distance operators their calculated
magnitudes are too small. 

Many authors \cite{NL} have recognized that $E$ and $A$ can have non-zero
phases relative to $T$ and $C$.  Our result, which distinguishes the present
analysis from the previous ones, is that $T$ and $C$ appear to differ from one
another in phase as well.  This phase difference appears to be similar (about
$150^\circ$) in $\bar K \pi$, $\bar K^* \pi$, and $\bar K \rho$ decays.  It
probably arises as a result from rescattering. The $E$ amplitude in $D^0$
decays and the $A$ amplitude in $D_s$ decays are found to have large phases
with respect to both $C$ and $T$.  We find that if the relative contributions
of $T$, $C$, and $E$ are such as to give large relative phases between
amplitudes for the three different charge states in $D \to \bar K \pi$ and $D
\to \bar K^* \pi$, then these phases naturally cancel in $D \to \bar K \rho$,
leaving amplitudes which are real with respect to one another. The $D \to \bar
K \rho$ amplitudes do contain contributions from final-state interactions, but
they are masked by cancelling phases.  This has important implications if one
wishes to ascribe final-state interactions to the proximity of resonances
\cite{Kamal,HJL}. 

This paper is organized as follows.  In Section II we review the flavor-SU(3)
decomposition of amplitudes and introduce notation for invariant amplitudes.
We tabulate the processes of interest, their decay rates, and their amplitudes
in Section III.  Then, in Section IV, we extract reduced amplitudes from the
data, and display pictorially their magnitudes and phases.  The amplitudes $T$
extracted in this way are compared with predictions from factorization and
semileptonic decays in Section V.  A brief discussion of resonant contributions
is contained in Section VI.  The role of disconnected diagrams involving $\eta$
and $\eta'$ production, which may be important in the decays $D_s \to \rho^+ +
(\eta,\eta')$, is discussed in Section VII, while Section VIII concludes. 
\newpage

\centerline{\bf II.  NOTATION}
\bigskip

Our meson wave functions are assumed to have the following quark content,
with phases chosen so that isospin multiplets contain no relative
signs \cite{GHLR,eta}:

\begin{itemize}

\item {\it Charmed mesons:}  $D^0 = - c \bar u$, $D^+ = c \bar d$, $D_s^+ =c
\bar s$. 

\item {\it Pseudoscalar mesons $P$:} $\pi^+ = u \bar d$, $\pi^0 = (d \bar d - u
\bar u)/\s$, $\pi^- = - d \bar u$, $K^+ = u \bar s$, $K^0 = d \bar s$, $\bar
K^0 = s \bar d$, $K^- = - s \bar u$, $\eta = (s \bar s - u \bar u - d \bar
d)/\st$, $\eta' = (u \bar u + d \bar d + 2 s \bar s)/\sx$.  (Here we adopt a
specific ansatz \cite{eta} for octet-singlet mixing in the $\eta$ and $\eta'$
wave functions.) 

\item {\it Vector mesons $V$:} $\rho^+ = u \bar d$, $\rho^0 = (d \bar d - u
\bar u)/\s$, $\rho^- = - d \bar u$, $\omega = (u \bar u + d \bar d)/\s$,
$K^{*+} = u \bar s$, $K^{*0} = d \bar s$, $\bar K^{*0} = s \bar d$, $K^{*-} = -
s \bar u$, $\phi = s \bar s$. 

\end{itemize}

The partial width $\Gamma$ for a specific two-body decay to $PP$ is expressed
in terms of an invariant amplitude ${\cal A}$ as 
\beq
\Gamma(D \to PP) = \frac{p^*}{8 \pi M^2}|{\cal A}|^2~~~,
\eeq
where $p^*$ is the center-of-mass (c.m.) 3-momentum of each final particle,
and $M$ is the mass of the decaying particle.  The kinematic factor of $p^*$ is
appropriate for the S-wave final state.  The amplitude ${\cal }$ will thus
have dimensions of (energy)$^{-1}$.

For $PV$ decays a P-wave kinematic factor is appropriate instead, and
\beq
\Gamma(D \to PV) = \frac{(p^*)^3}{8 \pi M^2}|{\cal A'}|^2~~~.
\eeq
Here ${\cal A'}$ is dimensionless.  These conventions agree with those of Chau
\ite~\cite{Chau}. 
\bigskip

\centerline{\bf III.  DECAY RATES AND AMPLITUDES}
\bigskip

In Tables I and II we summarize the rates, invariant amplitudes, and their
flavor-SU(3) representations for decays of charmed mesons to two pseudoscalar
mesons and to one pseudoscalar and one vector, respectively.  The branching
ratios are taken from the compilation of Ref.~\cite{PDG} except for branching
ratios for $D_s \to (\pi^+,\rho^+) + (\eta,\eta')$ from Ref.~\cite{Jessop}, and
are converted to decay rates using charmed particle lifetimes which are
averages \cite{JRFSI} of those in Ref.~\cite{PDG} and new CLEO values
\cite{CLEOlife}: $\tau(D^+) = 1051 \pm 31$ fs, $\tau(D^0) = 412.7 \pm 3.2$ fs,
$\tau(D_s^+) = 477 \pm 12$ fs. 

In Table I the amplitudes $T$, $C$, $E$, and $A$ were described above; in
Table II the amplitudes are labelled with subscripts which denote the meson
containing the spectator quark:  $P$ for pseudoscalar, $V$ for vector
\cite{DGR}.  

We omit contributions of disconnected diagrams \cite{ChauDisc,BFT} in which
$\eta$ and $\eta'$ exchange no quark lines with the rest of the diagram, and
couple through their SU(3)-singlet components.  Such diagrams are apparently
important for the understanding of the decays $B \to K \eta'$ \cite{DGRPRL}. 
They will be discussed in Sec.~VII. 

\begin{table}
\caption{Rates and invariant amplitudes for Cabibbo-favored decays of charmed
mesons to two pseudoscalar mesons.}
\begin{center}
\begin{tabular}{r c c c c c} \hline \hline
Decay & $M$   & Rate                 & $p^*$ & $|{\cal A}|$ & Representation \\
      & (GeV) & $(10^{10}{\rm s}^{-1})$ & (MeV) & $(10^{-6} {\rm GeV})$ \\
\hline \hline
$D^+ \to \bar K^0 \pi^+$& 1.8693 & $2.75 \pm 0.25$ & 862 & $1.36 \pm 0.06$
                        & $C + T$ \\ \hline
$D^0 \to K^- \pi^+$     & 1.8646 & $9.33 \pm 0.23$ & 861 & $2.50 \pm 0.03$
                        & $T + E$ \\
$ \to \bar K^0 \pi^0$   &        & $5.14 \pm 0.51$ & 860 & $1.85 \pm 0.09$
                        & $(C - E)/\s$ \\
$ \to \bar K^0 \eta$    &        & $1.72 \pm 0.24$ & 772 & $1.13 \pm 0.08$
                        & $C/\st$ \\
$ \to \bar K^0 \eta'$   &        & $4.17 \pm 0.63$ & 565 & $2.06 \pm 0.16$
                        & $-(C + 3E)/\sx$ \\ \hline
$D_s^+ \to \bar K^0 K^+$& 1.9685 & $7.54 \pm 3.20$ & 850 & $2.38 \pm 0.36$
                        & $C + A$ \\
$ \to \pi^+ \eta$       &        & $3.63 \pm 0.99$ & 902 & $1.61 \pm 0.22$
                        & $(T - 2A)/\st$ \\
$ \to \pi^+ \eta'$      &        & $7.78 \pm 2.15$ & 743 & $2.59 \pm 0.36$
                        & $[2(T + A)]/\sx$ \\ \hline \hline
\end{tabular}
\end{center}
\end{table}

\begin{table}
\caption{Rates and invariant amplitudes for Cabibbo-favored decays of charmed
mesons to one pseudoscalar and one vector meson.}
\begin{center}
\begin{tabular}{r c c c c c} \hline \hline
Decay & $M$   & Rate                & $p^*$ & $|{\cal A'}|$ & Representation \\
      & (GeV) & $(10^{10}{\rm s}^{-1})$ & (MeV) & $(10^{-6})$ \\
\hline \hline
$D^+ \to \bar K^{*0} \pi^+$& 1.8693 & $1.81 \pm 0.18$ & 712 & $1.70 \pm 0.09$
                           & $T_V + C_P$ \\ \hline
$D^+ \to \bar K^0 \rho^+$  &        & $6.28 \pm 2.38$ & 680 & $3.40 \pm 0.64$
                           & $T_P + C_V$ \\ \hline
$D^0 \to K^{*-} \pi^+$     & 1.8646 & $12.4 \pm 0.97$ & 711 & $4.45 \pm 0.17$
                           & $T_V + E_P$ \\
$ \to K^- \rho^+$          &        & $26.2 \pm 2.4$  & 678 & $6.95 \pm 0.32$
                           & $T_P + E_V$ \\
$ \to \bar K^{*0} \pi^0$   &        & $7.75 \pm 0.97$ & 709 & $3.54 \pm 0.22$
                           & $(C_P - E_P)/\s$ \\
$ \to \bar K \rho^0$       &        & $2.93 \pm 0.41$ & 676 & $2.34 \pm 0.16$
                           & $(C_V - E_V)/\s$ \\
$ \to \bar K^{*0} \eta$    &        & $4.60 \pm 1.21$ & 580 & $3.68 \pm 0.48$
                           & $(C_P + E_P - E_V)/\st$ \\
$ \to \bar K^{*0} \eta'$   &        & $ < 0.27$       &  99 & $ < 13$
                           & $-(C_P + E_P + 2 E_V)/\sx$ \\
$ \to \bar K^0 \omega$     &        & $5.09 \pm 0.97$ & 670 & $3.12 \pm 0.30$
                           & $-(C_V + E_V)/\s$ \\
$ \to \bar K^0 \phi$       &        & $2.08 \pm 0.24$ & 520 & $2.92 \pm 0.17$
                           & $-E_P$ \\ \hline
$D_s^+ \to \bar K^{*0} K^+$& 1.9685 & $6.91 \pm 1.89$ & 682 & $3.74 \pm 0.51$
                           & $C_P + A_V$ \\
$ \to \bar K^0 K^{*+}$     &        & $9.01 \pm 2.93$ & 683 & $4.26 \pm 0.69$
                           & $C_V + A_P$ \\
$ \to \rho^+ \eta$         &        & $22.5 \pm 6.5$  & 727 & $6.13 \pm 0.89$
                           & $(T_P - A_P - A_V)/\st$ \\
$ \to \rho^+ \eta'$        &        & $21.0 \pm 6.1$  & 470 & $11.4 \pm 1.7$
                           & $[2T_P + A_P + A_V)]/\sx$ \\
$ \to \pi^+ \rho^0$        &        & $< 0.17$        & 827 & $< 0.44$
                           & $(A_V - A_P)/\s$ \\
$ \to \pi^+ \omega$        &        & $0.65 \pm 0.29$ & 822 & $0.87 \pm 0.20$
                           & $(A_V + A_P)/\s$ \\
$ \to \pi^+ \phi$          &        & $7.54 \pm 1.89$ & 712 & $3.66 \pm 0.46$
                           & $T_V$  \\  \hline \hline
\end{tabular}
\end{center}
\end{table}

\newpage
\centerline{\bf IV.  REDUCED AMPLITUDES:  MAGNITUDES AND PHASES}
\bigskip

Rather than performing a $\chi^2$ fit, we show what information each amplitude
provides, and build up a graphical construction of the reduced amplitudes $T$,
$C$, $E$, etc., which exhibits their relative phases and magnitudes.  In this
way it is easier to spot regularities.  We are not greatly concerned with
errors on the fitted quantities in the present work, since most decays are
well-fitted while $D_s \to \rho^+\eta'$ is notably poorly reproduced, as has
been noted elsewhere \cite{ChauDisc,BFT,CT}.
\bigskip

\leftline{\bf A.  PP decays}
\bigskip

In the limit in which disconnected graphs do not contribute to $D \to PP$
decays, we find that $|C|$ is given by the $D^0 \to \bar K^0$ amplitude:
\beq
|C| = \sqrt{3}|{\cal A}(\bar K^0 \eta)| = (1.96 \pm 0.14) \times 10^{-6}~{\rm
GeV}~~~, 
\eeq
while by taking appropriate combinations of squares of amplitudes for
$D^0 \to \bar K^0 \pi^0$, $D^0 \to \bar K^0 \eta$, and $D^0 \to \bar K^0 \eta'$
we can eliminate the $C$--$E$ interference term to obtain
\beq
|E| = \left\{ \frac{1}{2}[|{\cal A}(\bar K^0 \pi^0)|^2 + |{\cal A}(\bar K^0
\eta')|^2] - |{\cal A}(\bar K^0 \eta)|^2 \right\}^{1/2} = (1.60 \pm 0.13)
\times 10^{-6}~{\rm GeV}~~~.
\eeq
The relative phase between $C$ and $E$ is given by
\beq
\cos \delta_{CE} = \left[ \frac{1}{4}|{\cal A}(\bar K^0 \eta')|^2 +
|{\cal A}(\bar K^0 \eta)|^2 - \frac{3}{4} [|{\cal A}(\bar K^0 \pi^0)|^2
\right]/|C||E| = - 0.07 \pm 0.11~~~,
\eeq
or $\delta_{CE} = (94 \pm 6)^\circ$.  The amplitudes $C$ and $E$ are depicted
in Fig.~1, along with a line $C-E = \sqrt{2}{\cal A}(D^0 \to \bar K^0 \pi^0)$.

\begin{figure}
\centerline{\epsfysize = 3 in \epsffile {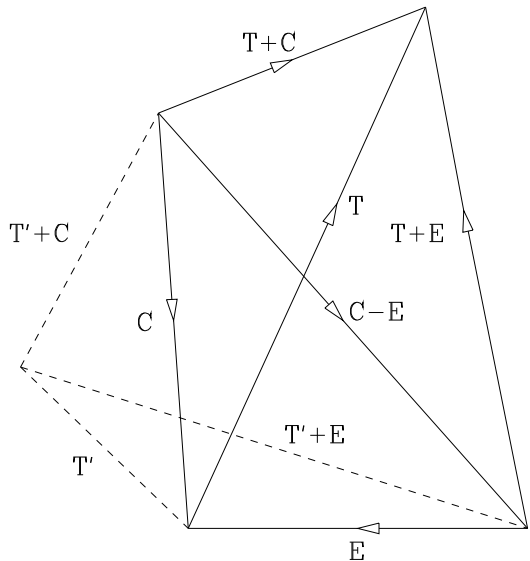}}
\caption{Amplitudes $T$, $C$, $E$ contributing to $D \to PP$ decays.
$T'$ denotes an alternate (disfavored) solution.}
\end{figure}

Next we use the rates for $D^+ \to \bar K^0 \pi^+$ and $D^0 \to K^- \pi^+$ to
specify the magnitudes of $T+C$ and $T+E$, respectively.  Lines corresponding
to these amplitudes form a triangle together with $C-E$ in the complex plane, a
consquence of the isospin relation ${\cal A}(K^- \pi^+) + \s{\cal A}(\bar K^0
\pi^0) = {\cal A}(\bar K^0 \pi^+)$.  This triangle can have either of two
orientations corresponding to reflection about the line corresponding to $C-E$.
These orientations correspond to different values of $T$. In Fig.~1 we denote
the favored orientation by solid lines.  With this choice, the value of $|T|
\simeq 2.7 \times 10^{-6}~{\rm GeV}$ is closer to that predicted by
factorization $|T| \simeq 2.0 \times 10^{-6}~{\rm GeV}$ (Sec.~V), and $|C| <
|T|$ as one might expect for a color-suppressed amplitude.  The other choice,
shown by the dashed lines and the primed amplitude $T'$, has $|C| > |T'| \simeq
1.1 \times 10^{-6}~{\rm GeV}$.  The determination of $T$ and $T'$ numerically
is a simple matter of solving a pair of simultaneous quadratic equations; the
central values are shown in Table III.  In what follows we shall consider only
the large-$|T|$ solution.

To test the above construction for consistency (particularly for the validity
of the assumption that no additional amplitudes are needed to describe decays
involving $\eta$ and $\eta'$) we consider the $D_s$ decays listed in Table I.
We can extract the magnitude of the ``annihilation'' amplitude $|A|$ from the
sum 
\beq
|{\cal A}(\pi^+ \eta)|^2 + |{\cal A}(\pi^+ \eta')|^2 = |T|^2 + 2 |A|^2
\eeq
and the value of $|T|^2$ from Table III to obtain $|A| = 1.01 \times
10^{-6}~{\rm GeV}$.  

Using the magnitudes of $T-2A$ and $T+A$ implied by Table I, we then may solve
for the phase of $A$ and the corresponding magnitude of $C+A$.  The two
solutions are shown in Fig.~2 and summarized in Table IV.  The solution
corresponding to the unprimed amplitude $A$ in Fig.~2 agrees with the value
$|C+A| = (2.38 \pm 0.36) \times 10^{-6}~{\rm GeV}$ implied in Table I by the
rate for $D_s^+ \to \bar K^0 K^+$, while that corresponding to the primed
amplitude $A'$ gives too small a value of $|C+A'|$. 

\begin{table}
\caption{Central values for the amplitudes $T$ (large solution) or $T'$ (small
solution) based on the decays $D \to (\bar K \pi, \bar K \eta, \bar K \eta')$.}
\begin{center}
\begin{tabular}{c c c c} \hline \hline
Solution & $|T|~(10^{-6}~{\rm GeV})$ & $|\delta_{ET}|$ & $|\delta_{CT}|$ \\
\hline
Large $|T|$ & 2.69 & $114^\circ$ & $152^\circ$ \\
Small $|T|$ & 1.08 & $ 44^\circ$ & $138^\circ$ \\ \hline \hline
\end{tabular}
\end{center}
\end{table}

Aside from an irrelevant sign, it is interesting that the phases of $E$ and
$A$ are almost identical.  This could be a sign of the universal behavior of
rescattering contributions conjectured in Ref.~\cite{JRFSI}.  The fact that
the magnitudes are not too different from one another is interesting, but
we do not have a ready explanation for it at the moment.

So far we have merely shown that there is a consistent solution for the
amplitudes in Cabibbo-favored decays of charmed mesons to $PP$.  The one
test of this consistency is the agreement of the predicted rate for
$D_s^+ \to \bar K^0 K^+$ when one of the discrete solutions for amplitudes is
chosen.  The comparison of this set of amplitudes with ones contributing to
Cabibbo-favored $PV$ decays, however, suggests that the solution may have
some validity.
\bigskip

\begin{figure}
\centerline{\epsfysize = 3 in \epsffile {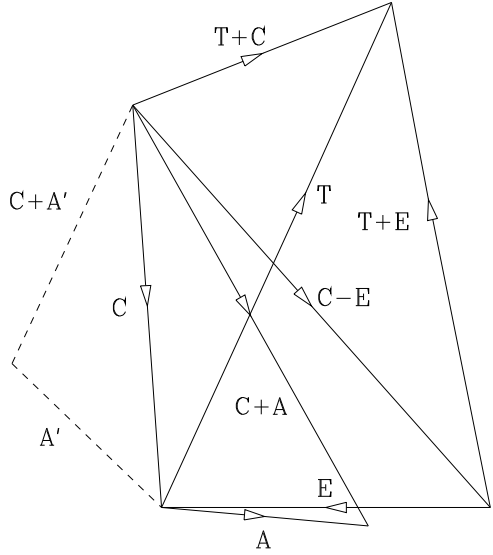}}
\caption{Amplitudes $T$, $C$, $E$, $A$ contributing to $D \to PP$ decays.
$A'$ denotes an alternate (disfavored) solution.}
\end{figure}

\begin{table}
\caption{Parameters of the solutions for $A$.}
\begin{center}
\begin{tabular}{c c c} \hline \hline
Solution & $|\delta_{AE}|$ & $|C+A|~(10^{-6}~{\rm GeV})$ \\
\hline
Favored    & $175^\circ$ & 2.35 \\
Disfavored & $ 36^\circ$ & 1.38 \\ \hline \hline
\end{tabular}
\end{center}
\end{table}

\leftline{\bf B.  PV decays}
\bigskip

The magnitudes of $T_V$ and $E_P$ are given by the $D_s \to \pi^+ \phi$ and
$D^0 \to \bar K^0 \phi$ amplitudes, respectively: 
$$
|T_V| = |{\cal A'}(\pi^+ \phi)| = (3.66 \pm 0.46) \times 10^{-6}~~~,
$$
\beq
|E_P| = |{\cal A'}(\bar K^0 \phi)| = (2.92 \pm 0.17) \times 10^{-6}~~~,
\eeq
The relative phase of $T_V$ and $E_P$ is given by
\beq
\cos(\delta_{E_P,T_V}) = \left[ |{\cal A'}(K^{*-}\pi^+)|^2 -
|{\cal A'}(\pi^+ \phi)|^2 - |{\cal A'}(\bar K^0 \phi)|^2 \right]
/2|T_V||E_P| = -0.10 \pm 0.18~~~,
\eeq
or $\delta_{E_P,T_V} = (96 \pm 10)^\circ$.  The amplitudes $T_V$ and $E_P$ are
shown in Fig.~3, along with the line $T_V+E_P = {\cal A'}((K^{*-}\pi^+)$.  We
neglect disconnected graphs involving $\omega$ and $\phi$ since, in contrast
to $\eta$ and $\eta'$, these seem to satisfy the Okubo-Zweig-Iizuka (OZI)
rule \cite{OZI} well in a wide variety of processes.

The rates for $D^+ \to \bar K^{*0} \pi^+$ and $D^0 \to \bar K^{*0} \pi^0$
then specify the magnitudes of $T_V + C_P$ and $C_P - E_P$, leading to two
possible solutions.  In one solution, $|C_P| < |T_V|$, as expected for a
color-suppressed amplitude, while in the other, $|C_P'| > |T_V|$, which we
regard as disfavored.  The favored solution is denoted by solid lines in
Fig.~3, while the disfavored solution is denoted by dashed lines.  The two
solutions are compared in Table V.  Note that for the favored solution, the
relative phase of the color-suppressed and tree amplitudes is exactly the same
($152^\circ$) as in the $D \to PP$ case analyzed above. 

\begin{figure}
\centerline{\epsfysize = 3 in \epsffile {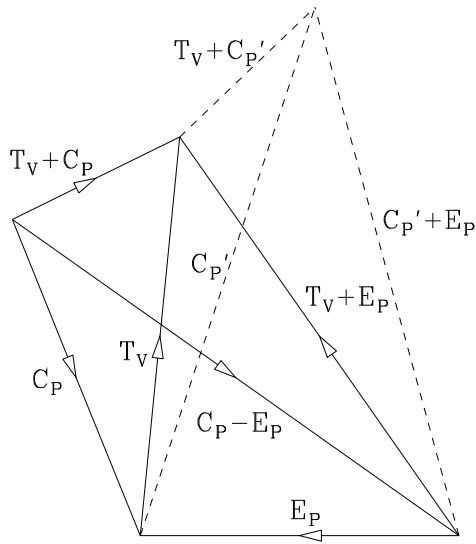}}
\caption{Amplitudes $T_V$, $C_P$, $E_P$ contributing to $D \to PV$ decays.
$C_P'$ denotes an alternate (disfavored) solution.}
\end{figure}

\begin{table}
\caption{Central values for the amplitudes $C_P$ (favored solution) or $C_P'$
(disfavored solution) based on the decays $D \to (\bar K^* \pi, \bar K^0 \phi)$
and $D_s^+ \to \pi^+ \phi$.}
\begin{center}
\begin{tabular}{c c c c} \hline \hline
Solution & $|C_P|~(10^{-6})$ & $|\delta_{C_P,E_P}|$ & $|\delta_{C_P,T_V}|$ \\
\hline
Favored    & 3.11 & $112^\circ$ & $152^\circ$ \\
Disfavored & 5.08 & $ 72^\circ$ & $168^\circ$ \\ \hline \hline
\end{tabular}
\end{center}
\end{table}

A further set of amplitudes may be specified if one is willing to assume that
$E_V = - E_P$.  This assumption is reasonable if the $E$ amplitude is dominated
by a quark-antiquark intermediate state, since it is then a consequence of
charge-conjugation invariance.  It is equivalent to the assumption made by
Lipkin \cite{HJLeta} in discussing the relative penguin contributions to $B \to
K^* \eta$ and $B \to K^* \eta'$ decays.  One can then construct a set of
amplitudes based on the decays $D \to (\bar K \rho, \bar K \omega, \bar K
\phi)$.

One first notes that the $D \to \bar K \rho$ amplitudes barely satisfy the
isospin triangle relation ${\cal A'}(\bar K^0 \rho^+) = {\cal A'}(K^-
\rho^+) + \sqrt{2} {\cal A'}(\bar K^0 \rho^0)$.  We perform a $\chi^2$ fit
in which the amplitudes are relatively real and find the best fit when
\beq
|{\cal A'}(\bar K^0 \rho^+)| = 3.57 \times 10^{-6}~~,~~~
|{\cal A'}(K^- \rho^+)|      = 6.90 \times 10^{-6}~~,~~~
|{\cal A'}(\bar K^0 \rho^0)| = 3.33 \times 10^{-6}~~~,
\eeq
which correspond to minor displacements from the central values in Table II.
We consequently shift those central values while maintaining the experimental
errors.

By combining the squared amplitudes for $D^0 \to K^0 \rho^0$ and $D^0 \to
\rho^0 \omega$ we then obtain $|E_V|^2 + |C_V|^2 = (15.28 \pm 2.01) \times
10^{-12}$, and recalling our assumption that $E_V = - E_P$, with $|E_P|^2
= (8.51 \pm 0.98) \times 10^{-12}$, we have
\beq
|C_V| = 2.60 \pm 0.43~~~.
\eeq
Furthermore,
\beq
\cos \delta_{C_V,E_V} = \left[ |{\cal A'}(\bar K^0 \omega)|^2 - |{\cal A'}(\bar
K^0 \rho^0)|^2 \right]/2|C_V||E_V| = 0.28 \pm 0.14~~~, 
\eeq
or $\delta_{C_V,E_V} = (74 \pm 8)^\circ$.  The amplitudes $E_V$, $C_V$, and
$C_V - E_V$ are shown in Fig.~4.

\begin{figure}
\centerline{\epsfysize = 3 in \epsffile {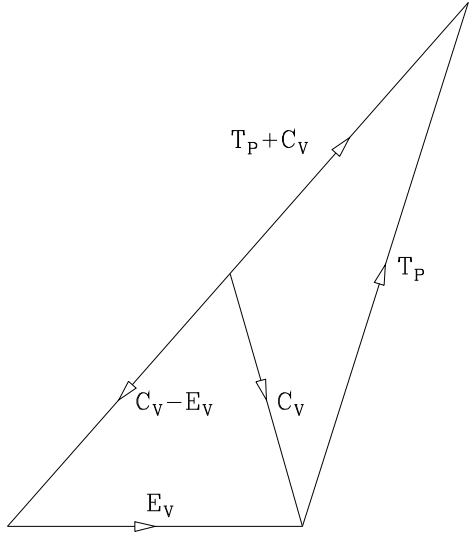}}
\caption{Amplitudes $T_P$, $C_V$, $E_V$ contributing to $D \to \bar K
(\rho,\omega,\phi)$ decays.}
\end{figure}

Since the best fit to $D \to \bar K \rho$ decays is obtained when all three
amplitudes are relatively real, we use the fact that ${\rm Arg}(T_P + E_V)
= {\rm Arg}(E_V - C_V) \simeq 49^\circ$ to construct the amplitudes
$T_P + E_V$, $T_P + C_V$, and $T_P$.  The results are:
\beq
|T_P| = 5.44 \times 10^{-6}~~,~~~\delta_{E_V,T_P} = 72^\circ~~,~~~
\delta_{C_V,T_P} = 148^\circ~~~.
\eeq
The relative phase of the color-suppressed and tree amplitudes is very similar
to that in the two previous constructions, as one sees from Fig.~4. 

The shape of Fig.~4 is very different from that of the two previous figures.
The fact that the $\bar K \rho$ amplitudes are in phase with one another
appears to be the consequence of the sign flip of $E_V$ relative to $E_P$ or
$E$.  This would not have had such a noticeable effect were it not for the fact
that, at least in our fits, the tree and color-suppressed amplitudes all
possess a relative phase of about $150^\circ$.

One more test of the assumption $E_V = - E_P$ is passed at about the $1 \sigma$
level.  The amplitude for $D^0 \to \bar K^{*0} \eta$ is predicted to be
${\cal A'}(\bar K^{*0} \eta) = (C_P + 2 E_P)/\st$ and, as a consequence of the
amplitudes determined above, is predicted to have magnitude $|{\cal A'}(\bar
K^{*0} \eta)| = 3.17 \times 10^{-6}$.  This is in satisfactory agreement
with the experimental value $|{\cal A'}(\bar K^{*0} \eta)| = (3.68 \pm 0.48)
\times 10^{-6}$.  The amplitude for $D^0 \to \bar K^{*0} \eta'$ is predicted to
be ${\cal A'}(\bar K^{*0} \eta) = -(C_P - E_P)/\sx = -{\cal A'}(\bar K^{*0}
\pi^0)/\st$ with magnitude $|{\cal A'}(\bar K^{*0} \eta)| = (2.04 \pm 0.13)
\times 10^{-6}$, much smaller than the current experimental upper bound.
\bigskip

\leftline{\bf C.  Comments on final-state interactions}
\bigskip

The conclusion of the fits to Cabibbo-favored $D \to PP$ and $D \to PV$
amplitudes is that final-state interactions (parametrized by large $E$ and $A$
contributions) are important in {\it all} final states, including the
$D \to \bar K \rho$ decays where the amplitudes for the three charge states
are all in phase with one another.  The presence of large final-state phases
in the $\bar K \rho$ case is masked by the cancellation of contributions
between the ``exchange'' amplitude and the ``color-suppressed'' amplitude.
This cancellation arises in $\bar K \rho$ decays and not in $\bar K^* \pi$
decays as a result of a sign flip in the ``exchange'' amplitude contribution,
which is just due to the charge-conjugation invariance of the strong coupling.
\bigskip

\centerline{\bf V.  FACTORIZATION COMPARISONS}
\bigskip

We compare the values of $T$, $T_V$, and $T_P$ obtained above with values
extracted using the factorization assumption \cite{BSW,NS,BJ,FM} and the
spectra $d \Gamma(D \to \bar K^{(*)} \ell^+ \nu_\ell)/ dq^2$, where $q =
p_{\ell} + p_{\nu_\ell}$.  For simplicity we use the limit of heavy-quark
effective theory and expressions derived in Ref.~\cite{FM}.  No QCD corrections
will be applied.  We shall neglect the pion mass. 
\bigskip

\leftline{\bf A.  $\bar K \pi$ decays}
\bigskip

We use the relation \cite{BJ}
\beq \label{eqn:BJ}
\frac{\Gamma(D \to \bar K \pi^+)_T}{d\Gamma(D \to \bar K \ell^+ \nu_\ell)/
dq^2|_{q^2 = m_\pi^2}} = 6 \pi^2 f_\pi^2 |V_{ud}|^2 = 0.98~{\rm GeV}^2~~~,
\eeq
where the subscript denotes the contribution of the $T$ amplitude to the $\bar
K \pi^+$ decay, excluding $C$ in $\bar K^0 \pi^+$ or $E$ in $K^- \pi^+$ (see
Table I).  Here $f_\pi = 132$ MeV.  A recent spectrum for $D^0 \to K^- \mu^+
\nu_\mu$ has been published by the Fermilab E687 Collaboration
~\cite{E687spec}.  Reading from their graph, we estimate 
\beq
\left. \frac{1}{\Gamma(D^0 \to K^- \mu^+ \nu_\mu)}\frac{d \Gamma(D^0 \to K^-
\mu^+ \nu_\mu)}{d q^2} \right|_{q^2 = m_\pi^2} = 0.76 \pm 0.09~{\rm GeV}^{-2}
~~~
\eeq
Furthermore, E687 quotes
\beq
\frac{\Gamma(D^0 \to K^- \mu^+ \nu_\mu)}{\Gamma(D^0 \to K^- \pi^+)}
= 0.852 \pm 0.034 \pm 0.028~~~.
\eeq
Putting these pieces together, we predict
\beq
\frac{\Gamma(D \to \bar K \pi^+)_T}{\Gamma(D^0 \to K^- \pi^+)} = 0.63 \pm 0.08
~~~,
\eeq
or $\Gamma(D \to \bar K \pi^+)_T = (5.9 \pm 0.8) \times 10^{10}~{\rm s}^{-1}$,
entailing $|T| = (1.99 \pm 0.13) \times 10^{-6}~{\rm GeV}$.  This is to be
compared with the result $|T| = 2.69 \times 10^{-6}~{\rm GeV}$ obtained in the
fit of Section IV:  $|T_{\rm fit}/ T_{\rm fact}| \simeq 1.35$.  This is well
within expectations of what QCD corrections might provide.  One must be careful
in applying such corrections in the present approach, however, since they will
mix operators of the type $T$, $C$, and $E$.  Our description is purely a
long-distance one.  A more complete treatment would probably involve a hybrid
between short- and long-distance effects. 
\bigskip

\leftline{\bf B.  $\bar K^* \pi$ decays}
\bigskip

In the heavy-quark limit, one expects $\Gamma(D \to \bar K^* \pi^+)_{T} =
\Gamma(D \to \bar K \pi^+)_{T}$ and near $q^2 = 0$ $d \Gamma(D \to \bar K^*
\ell^+ \nu_\ell)/dq^2 = d \Gamma(D \to \bar K \ell^+ \nu_\ell)/dq^2$.  In this
limit the $\bar K^*$ in the semileptonic process is longitudinally polarized.
Here we have used Eqs.~(9), (11), and (14) of Ref.~\cite{FM}.  Thus we predict
$\Gamma(D \to \bar K^* \pi^+)_T = (5.9 \pm 0.8) \times 10^{10}~{\rm s}^{-1}$, 
or, with the kinematic factors of Table I, $|T_V| = (3.08 \pm 0.21) \times
10^{-6}$.  Recall that $|T_{V,{\rm fit}}| = (3.66 \pm 0.46) \times 10^{-6}$,
where the error can be easily assigned since this quantity is obtained from the
decay $D_s \to \pi^+ \phi$.  Then $|T_{V,{\rm fit}}/T_{V,{\rm fact}}| = 1.19
\pm 0.17$. 
\bigskip

\leftline{\bf C.  $\bar K \rho$ decays}
\bigskip

A similar approach to $D \to \bar K \rho$ decays utilizes the semileptonic
spectrum of Fermilab E687 \cite{E687spec} with
\beq
\left. \frac{1}{\Gamma(D^0 \to K^- \mu^+ \nu_\mu)}\frac{d \Gamma(D^0 \to K^-
\mu^+ \nu_\mu)}{d q^2} \right|_{q^2 = m_\rho^2} = 0.74 \pm 0.09~{\rm GeV}^{-2}
\eeq
read from the graph, and $f_\pi \to f_\rho \simeq \sqrt{2} f_\pi$ \cite{FM}
in Eq.~(\ref{eqn:BJ}), with the result $\Gamma(D \to \bar K \rho)_T = (12.1 \pm
1.6) \times 10^{10}~{\rm s}^{-1}$, implying $|T_P| = (4.7 \pm 0.3) \times
10^{-6}$.  The fit in Section IV gave $|T_P| = 5.44 \times 10^{-6}$, so here
we have $|T_{P,{\rm fit}}/T_{P,{\rm fact}}| \simeq 1.15$.
\bigskip

\leftline{\bf D.  Summary of factorization results}
\bigskip

\begin{table}
\caption{Comparison of results for ``tree'' amplitudes obtained from fits to
data and using factorization.}
\begin{center}
\begin{tabular}{c c c c} \hline \hline
Method &        $|T|$        &    $|T_V|$    &   $|T_P|$   \\
       & ($10^{-6}~{\rm GeV}$) & ($10^{-6}$) & ($10^{-6}$) \\ \hline
Fit    &        2.69$^a$     & $3.66 \pm 0.46$ &  5.44$^a$ \\
Fact.  & $1.99 \pm 0.13$     & $3.08 \pm 0.21$ & $4.7 \pm 0.3$  \\
Fit/Fact. &    1.35$^a$      & $1.19 \pm 1.17$ & 1.15$^a$ \\ \hline \hline
\end{tabular}
\end{center}
\leftline{$^a$ Central value only.  Error on fitted amplitude was not
determined.}
\end{table}

We compare the results for $|T|$, $|T_V|$, and $|T_P|$ obtained from the fits
of Sec.~IV and those obtained via factorization in Table VI. All told, the
agreement with factorization for the ``tree'' amplitudes in $D \to \bar K \pi,
\bar K^* \pi, \bar K \rho$ decays is satisfactory.
\newpage

\centerline{\bf VI.  RESONANT INTERPRETATIONS}
\bigskip

We have found two relative phases in the present sets of fits:  those between
$C$ and $T$, and those between $E$ and $T$ (or $C$) amplitudes.  A resonant
interpretation of the $C$--$T$ relative phases is not possible; we ascribe
these phases rather to rescattering, most likely from the $T$ channel to the
$C$ channel rather than vice versa in view of the color-suppression of the weak
amplitude for the latter.  Thus, it will make most sense to examine the
relative phase between $E$ (or $A$) and $T$ amplitudes in terms of
contributions of possible direct-channel resonances. 
\bigskip

\leftline{\bf A.  PP decays}
\bigskip

A resonance contributing to the $E$ amplitude in $D \to \bar K \pi$ decays must
have spin-parity $J^P = 0^+$.  Such a resonance has been seen by the LASS
Collaboration \cite{LASS} with a mass of $M = 1945 \pm 10 \pm 20$ MeV/$c^2$ and
a width of $\Gamma = 201 \pm 34 \pm 79$ MeV.  [A reanalysis \cite{Anis} in a
$T$-matrix formalism quotes $M = 1820 \pm 40$ MeV/$c^2$ and $\Gamma = 250 \pm
100$ MeV.] 

In Fig.~1 and Table III we found the relative $T$--$E$ phase to be $114^\circ$.
This would indicate that $M_D$ was not far from a Breit-Wigner peak. If we
regard the $E$ amplitude as ``fed'' by rescattering from the elastic $K^-
\pi^+$ channel, we should take $T$ to be real and positive and $E$ to have a
positive imaginary part, in which case we should parametrize the propagator for
a resonance with mass $M_R$ and width $\Gamma_R$ as 
\beq \label{eqn:BW}
D(M) = \frac{1}{M_R - M - i \Gamma_R/2} = \frac{M_R - M + i \Gamma_R/2}
{(M - M_R)^2 + (\Gamma_R/2)^2}~~~,
\eeq
Then we would expect resonance dominance to give $\Gamma_R/[2(M_R - M_D)] =
\tan^{-1} \pm 114^\circ = - 2.3$.  This supports the claim \cite{Anis} that
$M_R < M_D$ and is compatible with the resonant parameters found in that
analysis. 

The $A$ amplitude seems to have a phase very close to that of $E$. No suitable
$I = 1$ resonance near $M_{D_s}$ with $J^P = 0^+$ appears in the most recent
compilation \cite{PDG}. 
\bigskip

\leftline{\bf B.  PV decays}
\bigskip

A resonance contributing to $D \to \bar K^* \pi$ or $D \to \bar K \rho$ decays
must have $J^P = 0^-$.  Normally one would expect such a resonance to have
equal and opposite couplings to $K^{*-} \pi^+$ and $K^- \rho^+$ channels,
by charge-conjugation invariance.  (In SU(3) language, one expects $F$-type
coupling of the resonance octet to the two final octets.)

The sign of the resonant contribution is less obvious in this case since we
expect rescattering to be fed by both the $K^{*-} \pi^+$ and $K^- \rho^+$
channels.  We use the result $|T_P| > |T_V|$, found both in our fits and in
a factorization calculation (cf. Table VI) to argue that the dominant
channel from which rescattering occurs is $K^- \rho^+$.  In that case it is
the $\bar K \rho$ channel for which the sign in Eq.~(\ref{eqn:BW}) applies.

In Sec.~IV B and Fig.~4, we found $|\delta_{E_V,T_P}| = 72^\circ$ for the $\bar
K \rho$ channels.  Since we expect $E_P = - E_V$ and no relative phase between
$T_P$ and $T_V$, we then predict $|\Delta_{E_P,T_V}| = \pi - |\delta_{E_V,T_P}|
= 108^\circ$, which is marginally consistent with the value $|\delta_{E_V,T_P}|
= (96 \pm 10)^\circ$ found in Sec.~IV B. 

One then expects $\Gamma_R/[2(M_R - M_D)] = \tan 72^\circ = 3.1$, or $M_R >
M_D$.  A $0^-$ resonance is seen \cite{Armstrong} in the vicinity of $M_D$, but
it is around 1830 MeV$/c^2$ and only its decay to $K^- \phi$ has been reported.
 If a resonant interpretation of the amplitudes $E_P$ and $E_V$ is correct, and
these amplitudes are generated mainly by rescattering from the dominant $K^-
\rho^+$ process, we expect there to exist a $0^-$ resonance slightly {\it
above} $M_D$, whose width should be about 6 times the $M_R - M_D$ difference,
decaying to $\bar K^* \pi$ and $\bar K \rho$, with equal partial widths aside
from small phase space corrections. 
\bigskip

\centerline{\bf VII.  AMPLITUDES NOT FITTED}
\bigskip

A group of amplitudes in Table II containing the contributions of $A_P$ and
$A_V$ has not been fitted.  As has been noted in Refs.~\cite{ChauDisc},
\cite{BFT}, and \cite{CT}, the decay $D_s \to \rho^+ \eta'$ cannot be fitted
without the introduction of additional contributions from disconnected diagrams
involving the flavor-SU(3)-singlet component of the $\eta'$, which will also
affect $\eta$ production to a small degree.  Such a component was anticipated
to be important in the decays $B \to K \eta'$ \cite{eta}, as has been borne out
experimentally \cite{CLEOetap} and widely discussed theoretically
\cite{etapth}.  As we have seen above, this component was not needed to fit any
of the other decays involving $\eta$ and $\eta'$, but its possible presence
could cast some doubt on the conclusions regarding the amplitude $A$ in $PP$
decays, as well as the parameters we have determined in Figs.~3 and 4 and Table
V.  Let us first recapitulate the tests for these parameters presented earlier.

The value of $A$ determined in Sec.~IV was found to be consistent with the
decay $D_s \to \bar K^0 K^+$.  The parameters of Fig.~3 and Table V were
found to be consistent with the rate for $D^0 \to \bar K^{*0} \eta$ and
with factorization.  (We do not count the prediction for the very small rate
for $D^0 \to \bar K^{*0} \eta'$ as much of a test since it relies mainly on the
very small available phase space.)  The parameters of Fig.~4 were consistent
with factorization.  It is possible that by appeal to Cabibbo-forbidden
decays and liberal use of (possibly broken) flavor-SU(3) one could glean
additional information, but that is beyond the scope of the present paper.

To see the nature of the problem, we compare
\beq
|{\cal A'}(\rho^+ \eta)|^2 + |{\cal A'}(\rho^+ \eta')|^2 = |T_P|^2
+ \frac{1}{2} [|A_P|^2 + |A_V|^2] = (167 \pm 39) \times 10^{-12}
\eeq
with
\beq \label{eqn:Abound}
0.42 \times 10^{-12} < |{\cal A'}(\pi^+ \rho^0)|^2 + |{\cal A'}(\pi^+
\omega)|^2 = |A_P|^2 + |A_V|^2 < 1.34 \times 10^{-12} 
\eeq
[using the upper bound on $\Gamma(\pi^+ \rho^0)$ and the $1 \sigma$ bounds
on $\Gamma(\pi^+ \omega)$] to conclude that $|T_P| > 11.3 \times 10^{-6}$, to
be compared with the fitted value of $5.4 \times 10^{-6}$.  At the same time,
if we omit the $\rho^+ \eta'$ decay from the fit, we find no difficulty in
constructing a set of amplitudes fitting the rates for $D_s \to (\bar K^{*0}
K^+, \bar K^0 K^{*+}, \pi^+ \omega)$ and the upper limit for $D_s \to \pi^+
\rho^0$, though the absence of a measurement for this last process prevents us
from specifying the parameters.  Since we already have information on $C_P$ and
$C_V$ (including their relative phase, which is small), we need both magnitudes
and phases for $A_P$ and $A_V$.  Without four measured decay rates or some
additional assumption, such information is unavailable. 

In view of the upper bound (\ref{eqn:Abound}) on the contribution of the
annihilation amplitudes, one might have expected the $\rho^+ \eta$ and $\rho^+
\eta'$ rates to be dominated by the $T_P$ amplitude.  However, comparing the
CLEO measurement \cite{Jessop} 
\beq
\frac{{\cal B}(D_s \to \rho^+ \eta')}{{\cal B}(D_s \to \rho^+ \eta)}
= 0.93 \pm 0.19
\eeq
with the corresponding ratio of semileptonic branching ratios \cite{Brand}
\beq
\frac{{\cal B}(D_s \to e^+ \nu_e \eta')}{{\cal B}(D_s \to e^+ \nu_e \eta)}
= 0.35 \pm 0.09 \pm 0.07~~~,
\eeq
there must be an additional contribution which is particularly important for
the decay $\rho^+ \eta'$.  Such a contribution would be provided by a
disconnected quark diagram. 

One might be tempted to ascribe $A_P$ and $A_V$ to the contribution of a $q
\bar q$ resonance.  However, such an interpretation would entail the relation
$A_P + A_V = 0$ (since there is no $I = 1$ $q \bar q$ $0^+$ resonance which can
couple to $\pi^+ \omega$).  This would run counter to the observation of $D_s
\to \pi^+ \omega$ and would entail very small values of both $A_P$ and $A_V$,
leading to difficulty in fitting the $\bar K^0 K^{*+}$ rate.  It is more likely
that $A_P$ and $A_V$ have a relative phase less than $\pi/2$ with respect to
each other and with respect to $C_P$ and $C_V$. 

We are left with the possibility that disconnected graphs play a role in the
decay $D_s \to \rho^+ \eta'$. The remaining processes seem to be described
satisfactorily without such contributions \cite{cf}, but some of them will be
affected when they are included.  In such a case, however, one cannot specify
the parameters of the fits without additional assumptions. 
\bigskip

\centerline{\bf VIII.  SUMMARY AND DISCUSSION}
\bigskip

The apparent puzzle of large relative phases between $D \to \bar K \pi$ and
$D \to \bar K^* \pi$ amplitudes but relatively real $D \to \bar K \rho$
amplitudes has been explained.  Amplitudes with large final-state phases are
present in all three classes of decays, but their effects are masked by
accidental cancellations in the $D \to \bar K \rho$ case.  The reason that this
cancellation can occur is that there are {\it two} types of amplitudes which
can have phases relative to the ``tree'' process $T$:  Both the
color-suppressed amplitudes $C$ and the exchange amplitudes $E$ (or
annihilation amplitudes $A$) have such phases.  

The relative phases between $C$ and $T$ amplitudes seem to be about $150^\circ$
in all three sets of processes.  These presumably arise from a rescattering
process in which the $C$ amplitudes are fed by $T$ contributions. A sign flip
in the $E$ amplitude is responsible for the difference between $\bar K^* \pi$
and $\bar K \rho$ behavior. 

The present fit implies tree amplitudes $T$ which are fairly close to those
obtained from semileptonic $D$ decays and factorization, and fits data for such
processes as $D_s \to \bar K^0 K^+$ and $D^0 \to \bar K^{*0} \eta$ without
additional parameters.  In company with a number of other approaches, it fails
to fit the decay $D_s \to \rho^+ \eta'$, since disconnected diagrams involving
the flavor-SU(3)-singlet component of the $\eta'$ have not been taken into
account.  These are expected to also play a (much smaller) role in the decay
$D_s \to \rho^+ \eta$, and possibly in other processes involving $\eta$ and
$\eta'$ as well. 

One might be tempted to draw conclusions about final-state effects in weak
decays of hadrons lighter or heavier than $D$'s from the above results.  In the
case of charmed mesons one sees that the amplitudes $A$ and $E$ are of
comparable magnitude to tree amplitudes $T$ and their color-suppressed versions
$C$.  This approximate equality is probably a way to understand why different
charmed particles differ in lifetimes by factors of a few, e.g.,
$\tau(D^+)/\tau(D^0) \simeq 2.5$. (Short-distance discussions of these
differences provide illuminating and probably complementary insights
\cite{BS}.)  In $K \to 2 \pi$ decays one can ascribe at least part of the
20-fold enhancement of the $I=0$ amplitude with respect to the $I=2$ amplitude
to such effects.  Thus, as the mass of the decaying particle increases by a
factor of 3, the effects of the final-state interactions seem to decrease (in
amplitude) by roughly a factor of 10. If this trend is extrapolated to $B$
particles, one would expect final-state-interaction amplitudes to be suppressed
with respect to ``tree'' processes by the same factor of 10.  If they have
large phases with respect to the tree processes they will not show up in
lifetime differences at present levels of sensitivity, while in certain cases
(e.g., $\Lambda_b$ decays) they might be in phase with tree amplitudes and
could give rise to effects in the 10--20\% range, perhaps accounting for the
observed ratio $\tau(\Lambda_b)/\tau(B) \simeq 0.8$ in contrast to less
successful attempts \cite{JRLB,MN} based purely on short-distance arguments. 

Final-state interaction effects also might lead to contributions interfering
with amplitudes such as the penguin amplitude assumed to dominate $B^+ \to \bar
K^0 \pi^+$ decays.  Elsewhere \cite{GHLR,BGR,GRresc} we have speculated that
amplitudes which involve the spectator quark (such as $E$ and $A$) would be
suppressed relative to amplitudes not involving the spectator (such as $T$) in
decays of mesons $M$ by a factor of $f_M/M_M$, where $f_M$ is the decay
constant of the corresponding meson.  This hierarchy does not appear to be
respected in the present example of charmed particles, where we find $|E| =
{\cal O}(|T|)$ but $f_D/M_D \simeq (200~{\rm MeV})/(1.9~{\rm GeV}) \simeq 0.1$.
 If, however, final-state effects fall off roughly as $1/M_M^2$ they will lead
to values of $|E/T|$ and $|A/T|$ closer to (but still in excess of) $f_B/M_B
\simeq (200~{\rm MeV})/(5~{\rm GeV}) \simeq 0.04$. 

The present description has been a purely long-distance one.  It could
probably be adapted to a hybrid treatment of both short- and long-distance
effects, reminiscent of that taken by Ciuchini \ite~\cite{Ciu} to describe
the enhancement of ``charming penguin'' amplitudes in $B$ decays.  Purely
short-distance descriptions of the matrix elements of penguin operators
fall short of those needed to explain a number of $B \to K \pi$ processes,
particularly in such processes as $B^+ \to K^+ \omega$ \cite{CLEOVP} in which
the spectator quark ends up in a vector meson \cite{DGR}.  The purely
short-distance approach to final-state phases \cite{BSS} involves calculations
of imaginary parts at the quark level, which could well underestimate the
importance of such effects even at masses as high as a few GeV/$c^2$.

\bigskip

\centerline{\bf ACKNOWLEDGEMENTS}
\bigskip

I would like to thank the Physics Department and the Theory Group at the
University of Hawaii for their hospitality during part of this work, and S.
Olsen, S. Pakvasa, S. F. Tuan, and H. Yamamoto for helpful conversations there.
I would also like to thank H. J. Lipkin for discussions on final-state
interactions. This work was supported in part by the United States Department
of Energy under Contract No. DE FG02 90ER40560. 
\bigskip

\def \ajp#1#2#3{Am.~J.~Phys.~{\bf#1}, #2 (#3)}
\def \apny#1#2#3{Ann.~Phys.~(N.Y.) {\bf#1}, #2 (#3)}
\def \app#1#2#3{Acta Phys.~Polonica {\bf#1}, #2 (#3)}
\def \arnps#1#2#3{Ann.~Rev.~Nucl.~Part.~Sci.~{\bf#1}, #2 (#3)}
\def \art{and references therein}
\def \b97{{\it Beauty '97}, Proceedings of the Fifth International
Workshop on $B$-Physics at Hadron Machines, Los Angeles, October 13--17,
1997, edited by P. Schlein}
\def \carg{{\it Masses of Fundamental Particles -- Carg\`ese 1996}, edited by
M. L\'evy \ite, NATO ASI Series B:  Physics Vol.~363 (Plenum, New York, 1997)}
\def \cmp#1#2#3{Commun.~Math.~Phys.~{\bf#1}, #2 (#3)}
\def \cmts#1#2#3{Comments on Nucl.~Part.~Phys.~{\bf#1}, #2 (#3)}
\def \corn93{{\it Lepton and Photon Interactions:  XVI International
Symposium, Ithaca, NY August 1993}, AIP Conference Proceedings No.~302,
ed.~by P. Drell and D. Rubin (AIP, New York, 1994)}
\def \cp89{{\it CP Violation,} edited by C. Jarlskog (World Scientific,
Singapore, 1989)}
\def \dpff{{\it The Fermilab Meeting -- DPF 92} (7th Meeting of the
American Physical Society Division of Particles and Fields), 10--14
November 1992, ed. by C. H. Albright \ite~(World Scientific, Singapore,
1993)}
\def \dpf94{DPF 94 Meeting, Albuquerque, NM, Aug.~2--6, 1994}
\def \efi{Enrico Fermi Institute Report No. EFI}
\def \el#1#2#3{Europhys.~Lett.~{\bf#1}, #2 (#3)}
\def \epjc#1#2#3{Eur.~Phys.~J.~C {\bf#1}, #2 (#3)}
\def \f79{{\it Proceedings of the 1979 International Symposium on Lepton
and Photon Interactions at High Energies,} Fermilab, August 23-29, 1979,
ed.~by T. B. W. Kirk and H. D. I. Abarbanel (Fermi National Accelerator
Laboratory, Batavia, IL, 1979}
\def \hb87{{\it Proceeding of the 1987 International Symposium on Lepton
and Photon Interactions at High Energies,} Hamburg, 1987, ed.~by W. Bartel
and R. R\"uckl (Nucl. Phys. B, Proc. Suppl., vol. 3) (North-Holland,
Amsterdam, 1988)}
\def \ib{{\it ibid.}~}
\def \ibj#1#2#3{{\it ibid.}~{\bf#1}, #2 (#3)}
\def \ichep72{{\it Proceedings of the XVI International Conference on High
Energy Physics}, Chicago and Batavia, Illinois, Sept. 6--13, 1972,
edited by J. D. Jackson, A. Roberts, and R. Donaldson (Fermilab, Batavia,
IL, 1972)}
\def \ijmpa#1#2#3{Int.~J.~Mod.~Phys.~A {\bf#1}, #2 (#3)}
\def \ite{{\it et al.}}
\def \jmp#1#2#3{J.~Math.~Phys.~{\bf#1}, #2 (#3)}
\def \jpg#1#2#3{J.~Phys.~G {\bf#1}, #2 (#3)}
\def \lkl87{{\it Selected Topics in Electroweak Interactions} (Proceedings
of the Second Lake Louise Institute on New Frontiers in Particle Physics,
15--21 February, 1987), edited by J. M. Cameron \ite~(World Scientific,
Singapore, 1987)}
\def \KEK#1{{\it Flavor Physics} (Proceedings of the Fourth International
Conference on Flavor Physics, KEK, Tsukuba, Japan, 29--31 October 1996),
edited by Y. Kuno and M. M. Nojiri, Nucl.~Phys.~B Proc.~Suppl.~{\bf 59},
#1 (1997)}
\def \ky85{{\it Proceedings of the International Symposium on Lepton and
Photon Interactions at High Energy,} Kyoto, Aug.~19-24, 1985, edited by M.
Konuma and K. Takahashi (Kyoto Univ., Kyoto, 1985)}
\def \mpla#1#2#3{Mod.~Phys.~Lett.~A {\bf#1}, #2 (#3)}
\def \nc#1#2#3{Nuovo Cim.~{\bf#1}, #2 (#3)}
\def \nima#1#2#3{Nucl.~Instr.~Meth.~A {\bf#1}, #2 (#3)}
\def \np#1#2#3{Nucl.~Phys.~{\bf#1}, #2 (#3)}
\def \npbps#1#2#3{Nucl.~Phys.~B (Proc.~Suppl.) {\bf#1}, #2 (#3)}
\def \pisma#1#2#3#4{Pis'ma Zh.~Eksp.~Teor.~Fiz.~{\bf#1}, #2 (#3) [JETP
Lett. {\bf#1}, #4 (#3)]}
\def \pl#1#2#3{Phys.~Lett.~{\bf#1}, #2 (#3)}
\def \plb#1#2#3{Phys.~Lett.~B {\bf#1}, #2 (#3)}
\def \pr#1#2#3{Phys.~Rev.~{\bf#1}, #2 (#3)}
\def \pra#1#2#3{Phys.~Rev.~A {\bf#1}, #2 (#3)}
\def \prd#1#2#3{Phys.~Rev.~D {\bf#1}, #2 (#3)}
\def \prl#1#2#3{Phys.~Rev.~Lett.~{\bf#1}, #2 (#3)}
\def \prp#1#2#3{Phys.~Rep.~{\bf#1}, #2 (#3)}
\def \ptp#1#2#3{Prog.~Theor.~Phys.~{\bf#1}, #2 (#3)}
\def \rmp#1#2#3{Rev.~Mod.~Phys.~{\bf#1}, #2 (#3)}
\def \rp#1{~~~~~\ldots\ldots{\rm rp~}{#1}~~~~~}
\def \si90{25th International Conference on High Energy Physics, Singapore,
Aug. 2-8, 1990}
\def \slc87{{\it Proceedings of the Salt Lake City Meeting} (Division of
Particles and Fields, American Physical Society, Salt Lake City, Utah,
1987), ed.~by C. DeTar and J. S. Ball (World Scientific, Singapore, 1987)}
\def \slac89{{\it Proceedings of the XIVth International Symposium on
Lepton and Photon Interactions,} Stanford, California, 1989, edited by M.
Riordan (World Scientific, Singapore, 1990)}
\def \smass82{{\it Proceedings of the 1982 DPF Summer Study on Elementary
Particle Physics and Future Facilities}, Snowmass, Colorado, edited by R.
Donaldson, R. Gustafson, and F. Paige (World Scientific, Singapore, 1982)}
\def \smass90{{\it Research Directions for the Decade} (Proceedings of the
1990 Summer Study on High Energy Physics, June 25 -- July 13, Snowmass,
Colorado), edited by E. L. Berger (World Scientific, Singapore, 1992)}
\def \stone{{\it B Decays}, edited by S. Stone (World Scientific,
Singapore, 1994)}
\def \tasi90{{\it Testing the Standard Model} (Proceedings of the 1990
Theoretical Advanced Study Institute in Elementary Particle Physics,
Boulder, Colorado, 3--27 June, 1990), edited by M. Cveti\v{c} and P.
Langacker (World Scientific, Singapore, 1991)}
\def \vanc{29th International Conference on High Energy Physics, Vancouver,
23--31 July 1998}
\def \yaf#1#2#3#4{Yad.~Fiz.~{\bf#1}, #2 (#3) [Sov.~J.~Nucl.~Phys.~{\bf #1},
#4 (#3)]}
\def \zhetf#1#2#3#4#5#6{Zh.~Eksp.~Teor.~Fiz.~{\bf #1}, #2 (#3) [Sov.~Phys.
-- JETP {\bf #4}, #5 (#6)]}
\def \zpc#1#2#3{Zeit.~Phys.~C {\bf#1}, #2 (#3)}


\begin{thebibliography}{99}
\bibitem{cf} In particular, we find no difficulty in fitting the decay
$D_s \to \pi^+ \eta'$ without disconnected diagrams, in contrast to the
conclusions of Refs.~\cite{ChauDisc} and \cite{BFT}.  However, the likelihood
that this process does receive contributions from such diagrams indicates that
the agreement we obtained with the observed rate for $D_s \to \bar K^0 K^+$
may be fortuitous.
\end{thebibliography}
\end{document}